# The Athermal Laser

## Peter Muys

Lambda Research Optics Europe N.V., Tulpenstraat 2, 9810 Eke, Belgium

pmuys@lambdaeurope.be

Abstract

A new laser concept is presented, called the athermal laser, unifying all the hitherto known implementations of radiative laser cooling. This cooling principle is based on counter-acting the heat generation in the laser's active material by triggering a radiative process which conveys away this heat. When spontaneous anti-Stokes scattering is considered for this transport, it is realised through three-wave interaction in the dopant (pump field, laser field and anti-Stokes field). It is shown here that by switching over to a four-wave mixing process in the dopant/host, this limitation can be overcome and the cooling radiation now can become stimulated, with a potentially higher cooling efficiency as a consequence. The athermal laser concept now encompasses spontaneous and stimulated radiative cooling and is found to remain in line with the classical thermodynamic picture of the laser as a combined heating/cooling machine. Moreover, we demonstrate that it automatically fulfils the radiation balancing condition between the heating and cooling stage. This lifts the highly stringent requirement of suitably overlapping emission and absorption spectra of the dopant in a quasi-two-level system, as is required for spontaneous radiative cooling. Finally, the balancing condition of the athermal laser translates naturally into a radiation temperature condition.

The balancing condition further implies that the stimulated cooling concept is not limited to dopants with a small quantum defect, as is the case for spontaneous cooling. This lifting opens up many new device applications, with a wide choice of laser materials, and includes Raman lasers, vibronic lasers and quasi-two-level lasers. Last but not least, it points the way to new cooling techniques for existing high power lasers and enables new resonator concepts.

# 1. Introduction

Cooling of doped solids by spontaneous anti-Stokes fluorescence is currently being studied intensively to realize a refrigerator which is only powered by a laser beam[1]. In parallel, efforts are made to apply the same principle to cool high power solid-state lasers, where thermal effects in the active medium are known to be detrimental to obtain high quality beams. In this way, radiation balanced lasers were conceived [2], where the spontaneous fluorescence compensates for the quantum defect heating. This picture was extended recently to stimulated radiative processes in vibronic [ ] and Raman lasers[ ], which have the potential to be much more energetic that their spontaneous counterpart.

In this paper, we will show how the concepts of spontaneous and stimulated cooling can be treated from a unified, thermodynamic point of view.

# 2. Cooling by spontaneous radiation

We start with the thermodynamic approach to laser devices [3] and consider the block diagrams of a regular optically pumped laser and of an optically pumped refrigerator. The laser-pumped laser is conceptually well known. The refrigerator on the other hand is still a bit of an exotic item. It consists of an optically active medium that resonantly absorbs all of the input laser radiation, see fig (1a). The thermally assisted anti- Stokes fluorescence of the dopant carries away the energy of the external cooling load, resulting in a chilling of this load. In fact, an optical refrigerator can be considered as a regular laser, but working in reverse, see Fig. (1b), since it transforms "high quality" laser photons in "low quality" spontaneous emission.

Both concepts, the laser as a heating machine and the refrigerator as a cooling machine, can be combined in one system, by connecting the load of the refrigerator to the laser, and then pump both simultaneously (by the same pump laser), as represented in Fig. (2). All the waste heat Q generated by the laser is removed by the refrigerator. This is called "radiation balancing" [2]. In this way the original two-stage series connection can be condensed to a single stage and can be practically implemented by using the same active material for both devices. Based on the condition for radiation balancing and on the steady-state form of the rate equations, it turns out [2] that the pump frequency $\omega_p$ must be chosen to be intermediate between the mean spontaneous emission frequency $\omega_a$ and the stimulated emission frequency $\omega_l$, according to:

$$\omega_l < \omega_p < \omega_a \qquad (1)$$

We will refer to this inequality as "the balancing condition". However innocent this condition might look, it imposes a stringent and non-trivial requirement on how the emission and absorption spectrum of the dopant overlap since $\omega_a$ is an averaged value over the spontaneous emission spectrum and hence depends critically on the spectrum's fine structure. If the balancing condition can be fulfilled, then in effect, one is balancing the anti-Stokes cooling from the refrigerator against the Stokes heating from the laser, resulting in athermal or radiation-balanced lasing. Because of the extremely narrow scope imposed by the radiation balance condition (1) on the spectra, it turns out in practice that only a few materials could be actually identified as exhibiting this property.

The dopant of the radiation balanced laser exhibits four energy levels, as depicted in fig. (3a), the ground level, the upper and the lower laser level and an excited level lying above the upper laser level which we call the fluorescence level. The upper level and the fluorescence level together belong to one band of levels. The ground and the lower level belong to a lower lying band. A rate equation analysis can be set up to describe the band-to-band transitions, relying on the theory of McCumber [4] and which ultimately leads to the balancing condition eq (1) [2]. The denomination [2] for this laser system is a quasi-two-level laser, a quasi-level being called here now a band. In fact , the spontaneous anti-Stokes emission turns out to be rather modest, imposed by the stringent balancing condition (1), and so this results in a limited cooling capability by this radiation channel. As a consequence, only dopants having a small quantum defect (typically 4% in practice) can be considered for this type of cooling.

### 3. Cooling by stimulated radiation

In ref.[5], the second law of thermodynamics and the concept of entropy is invoked to argue that a stimulated anti-Stokes process is not possible in a three-wave interaction. In ref.[6], it is further demonstrated that, using the concept of radiation entropy carried by an optical beam, the spontaneous emission process in a three-wave interaction shows to be orders of magnitude more efficient as cooling process than any other conceived coherent emission process. The proof is constructive and relies on calculating the cooling efficiency as a function of the radiative emission entropy of the output field. Stated somewhat differently, we interpret this result as a confirmation that stimulated anti-Stokes cooling during a three-wave interaction is not possible. Op het eerste zicht betekent dit dus dat het spoor om gestimuleerd te koelen hier doodloopt. En dat dus onze theorie nie just is. At first sight, this conjecture

In fact, the same negative conclusion can be reached, without invoking entropy arguments, by solving the equation of motion for an elementary molecular oscillator, according to the Lorentz model, for three-wave interaction in a Raman medium[7].The Lorentz model assumes that the vibrational mode of this medium is described as a simple harmonic oscillator with resonance frequency $\omega_v$ and damping constant $\gamma$. X(t) denotes the deviation of the internuclear distance from its equilibrium value and is determined by

$$\frac{d^2X}{dt^2} + \gamma \frac{dX}{dt} + \omega_v^2 X = \frac{\varepsilon_0}{2m}(\frac{\partial \alpha}{\partial x})_0 \langle E^2(t) \rangle \qquad (2)$$

$(\partial \alpha / \partial x)_0$ is derived from the polarizibility $\alpha$ of the oscillating Raman-active atom. The triangular brackets in the right-hand side of eq. (2) indicate an average over one optical cycle. The vibration is driven by the total applied optical field (pump + Stokes +anti-Stokes)

$$\tilde{E}(z,t) = A_p \exp[i(k_p z - \omega_p t)] + A_s \exp[i(k_s z - \omega_s t)] + A_a \exp[i(k_a z - \omega_a t)] + c.c. \qquad (3)$$

where $\omega_p + \omega_v = \omega_a$. Substituting eq. (3) in eq. (2) and looking for time-harmonic solutions of X(t), the total polarization P can then be determined as

$$P = \varepsilon_0 N \alpha(z,t) E(z,t)$$

where its nonlinear part is given by

$$P_{NL} = a\{bE_a E_p^* \exp[i(\omega_a - \omega_p)] + c.c.\}[E_p \exp(i\omega_p) + E_s \exp(i\omega_s) + c.c.] \qquad (4)$$

a and b are used as abbreviations for complicated coefficients, which are of no concern here in this context. This nonlinear polarization contains a term oscillating at $\omega_a$ and defining the third order susceptibility $\chi_a^{(3)}$

$$P_{NL}^{(\omega_a)} = \chi_a^{(3)} |E_p|^2 E_a \qquad (5)$$

the gain of this wave is given by

$$g_a = -\frac{k_s}{2n_a^3}|E_p|^2 \chi_a^{(3)} \qquad (6)$$

For this case, the anti-Stokes gain comes out negative, because the third-order susceptibility shows to be positive, and so the anti-Stokes wave attenuates. This fact is well known in the literature. But in the context of this paper, it can be interpreted in a new way and it effectively shows the impossibility of stimulated anti-Stokes scattering for this type of *three*-wave interaction.

Ref. [5] notes that alternatives might be found to work around this limitation: "another possible way of achieving laser emission in a nominally anti-Stokes process is when other processes are coupled to the medium so that its overall entropy change is positive". This means that in order to use the anti-Stokes scattering phenomenon nevertheless, we should consider a more complex interaction. In this paper we follow this strategy and propose a four-level system with a *four*-wave mixing process in the nonlinear medium with third-order susceptibility $\chi^{(3)}$.

Indeed, on closer inspection, it turns out that eq.(4) contains another term oscillating at $\omega_a$. Now, the driving term for the nonlinear polarization becomes proportional to $E_p E_p E_s^*$ and

$$\omega_a - \omega_p = \omega_p - \omega_s \qquad (7)$$

is required. Two pump photons interact with one Stokes and one anti-Stokes photon, which constitutes a particular form of four-wave mixing. The non-linear polarization stays an explicit function of time, according to

$$P_{NL}^{(\omega_a)} \propto E_p^2 E_s^* \exp[i(2\omega_p - \omega_s)t] \qquad (8)$$

so that the (vectorial) phase matching condition

$$2\mathbf{k}_p - \mathbf{k}_s = \mathbf{k}_a \qquad (9)$$

must be introduced.

As a direct and important consequence, the conclusion then follows for a Raman-active medium that a *positive* anti-Stokes gain can be realized so that the wave mixing now can become stimulated.

## 4. The doped non-linear host

To elaborate further on this conclusion, we consider a hybrid form of a laser-active medium and a Raman-active medium [8]. In particular, we consider the embedding of a laser-active dopant in a non-linear host, which is typically a solid-state material, see Fig. (3b). Four levels are involved: the common ground level g, the upper laser level u of the dopant, the terminal laser level t of the dopant and a virtual level of the host. The laser constitutes a vibronic or phonon-terminated laser, meaning that the t-level of the dopant is resonantly coupled to a vibrational level v of the host.

In order for the four-wave mixing process to take place with macroscopic efficiency, the host must show a high third-order nonlinear susceptibility $\chi^{(3)}$ and must show lattice oscillations which are strongly coupled to the lower level electronic t-states of the dopant. The lower level excited electronic states of the dopant hence form one common energy reservoir with the phonons of the host. These phonons are ready to participate in an anti-Stokes Raman scattering process, triggered by the pump photons and emptying in this way the vibrational level by emission of anti-Stokes photons.

The above situation describes a strongly non-equilibrium situation for the terminal laser level. The four-wave mixing process is depleting the phonon level so efficiently that the host tries to reach thermal equilibrium again by filling the vibrational level through extraction of phonon energy from the lattice. This actually means that energy is conveyed away from the crystal as anti-Stokes radiation and that in this way it is being cooled radiatively. As a consequence, the steady-state value for the occupational density of the vibrational level will be *lower* than its equilibrium value at that temperature.

The occurrence of phonon annihilation and the accompanying temperature decrease can be mathematically deduced from the equation of motion for the molecular vibration in the Lorentz model of the host alone [9]. It is not required to simultaneously solve the photon field equations. The differential equation (2) for the mechanical vibration is therefore solved by plugging in the test solution

$$X(z,t) = X(\Omega)\exp[i(Kz - \Omega t)] + c.c. \qquad (10)$$

where $k_p - k_s = k_a - k_p \equiv K$ because of the phase matching condition (9).

## 5. The radiation entropy balance during FWM

For radiation balancing in a quasi-two-level laser by spontaneous anti-Stokes fluorescence, the entropy balance as imposed by the second law of thermodynamics is rather easy to fulfill, due to the large entropy flux of the "disordered" spontaneous radiation. In case of stimulated cooling however, the overall entropy increase will be small, because of the "ordered" nature of the input and output beams. This increase is possible nevertheless, due to the conical shape of the escaping Stokes and anti-Stokes radiation, which are effectively radiating as a Bessel beams [10], and will consequently have a larger divergence than a Gaussian beam. The radiation entropy of this beam now is proportional to the beam divergence[11]. We can conclude that based on the second law of thermodynamics, stimulated radiative cooling is not forbidden.

## 6. The radiation energy balance during FWM

To analyze the temperature evolution of the ensemble, a rate equation model must be set up, describing the interaction dynamics of the different light fields with the population of the electronic levels of the dopant and with the total internal energy of the host. In most regular treatments of Raman scattering, the disturbance of the ground level by the scattering process is neglected , since one is more interested in the generated new optical fields than in the excitation state of the Raman medium. In our model however, the interaction with the dopant levels is crucial. The host internal energy is stored in the vibrational modes of the lattice and is given by

$$E = mc_v T \qquad (14)$$

We suppose the crystal thermally isolated from its environment, so that only adiabatic changes of its internal energy can take place. The contribution of the dopant is described by the rate equations for the upper $N_u$ and the terminal $N_t$ laser level. The upper level is filled by pumping out of the ground state reservoir, and is emptied by stimulated radiation and eventually by nonradiative decay which creates phonons in the host.

For cooling applications it is crucial that the interaction between all photon fields and all population densities are taken into account, together with the host internal energy. Photon densities for the fields are: $\Phi_p$ for the pump field, $\Phi_l$ for the laser field and $\Phi_a$ for the anti-Stokes field. The rate equation of the laser field takes into account that light is re-circulating in the cavity, by defining the cavity photon lifetime $\tau_c$. The fluorescence time of the anti-Stokes radiation is denoted as $\tau_a$. Finally, the evolution of the internal energy E is described by

$$\frac{dE}{dt} = c\,\sigma_p(N_g - N_u)G_p\Phi_p - \frac{\Phi_l}{\tau_c} + G_t\frac{N_t - N_{teq}}{\tau_t} - \frac{h\nu_a}{\tau_a}\Phi_a \qquad (15)$$

where $\sigma_p$ is the cross section of the pump transition, $G_p$ is the pump quantum energy and $G_t$ is the phonon quantum energy. $N_g$ and $N_u$ are the population densities of resp. the ground and upper laser level. This equation explicitly shows the energy input by absorption of the pump photons, the outcoupled laser light, the non-radiative contributions by the vibrational level to the phonon reservoir of the internal energy, and the draining of the generated heat carried away by the anti-Stokes beam. In a regularly cooled laser, this last term is replaced by the power which is carried away by thermal conduction into the heat sink. The third contribution to eq (15) should also be negative under a far-from-equilibrium situation, in order to be effective as a cooling term. In steady-state, the time derivative of the internal energy gets zero and eq (15) reduces to the extended radiation balance: the absorbed power density equals the radiated power density. Compared to the balance for spontaneous radiation, an extra term now shows up.

## 7. The general athermal laser concept

The preceding paragraphs have shown that spontaneous and stimulated cooling have may points of overlap. We will elaborate on these common aspects here and show that the balancing condition appears as a general property of the athermal laser concept. We therefore identify three generic types of lasers, based on the block diagrams of fig(1). The dopant and the host will be either a laser or a refrigerator, as given in fig(4). Three combinations are relevant for the concept. First, radiation balancing by spontaneous anti Stokes radiation occurs when the dopant is the laser and the dopant is also the refrigerator. Next, stimulated cooling occurs when the dopant is the laser and the host is the refrigerator. A third form is possible in a Raman laser, where the host constitutes the laser and the refrigerator. The corresponding energy levels are represented in fig (5).

The radiation balancing condition eq(1) for a quasi-two-level laser exhibiting spontaneous radiative cooling is a consequence of the particular and intricate overlap of the absorption and emission spectrum of the dopant. For stimulated emission however, the details of this overlap become irrelevant. Indeed, the balancing condition now follows from substituting the condition $\omega_p > \omega_l$ in eq.(7), leading to

$$\omega_a > \omega_p$$

which is again eq.(1). In other words, *stimulated cooling automatically fulfills the radiation balancing condition*, irrespective of the details of the band overlap structure of the dopant.

Finally, the conceptual links between spontaneous and stimulated radiative cooling are further detailed. A quasi-two-level, radiation balanced laser for spontaneous radiation requires the energy spectrum of the dopant to exhibit a band structure. In particular the lowest band must be connected to the phonon levels of the host material. This effectively means that we are looking here to a vibronic laser, as is the case for stimulated radiation balancing. In other words, given that the host is sufficiently nonlinear, this quasi-two-level laser could also be cooled by stimulated radiation.

## 8. Conclusions

In conclusion, we have shown that stimulated radiative laser cooling is thermodynamically not forbidden by the second law and is compatible with the principle of radiation balancing. The unified form of the radiation balance was derived, encompassing quasi-two-level lasers, vibronic lasers and Raman lasers. Even materials with a high quantum defect now can become candidate for stimulated cooling. The materials no longer need to show an intricate overlap between absorption and fluorescence spectrum, as was required for spontaneous cooling. This puts the door wide open for radiative cooling a whole new class of lasers, including Raman lasers .

# Figure captions

Fig 1a thermodynamic block diagram for an optically pumper laser
    1b thermodynamic block diagram for an optically pumped refrigerator

Fig 2 block diagram for the cooled laser

Fig 3a energy levels and transitions for spontaneous radiative laser cooling
    3b energy levels and transitions for stimulated radiative laser cooling

Fig 4: the unified approach of the athermal laser
    4a cooling in a dopant/dopant configuration
    4b cooling in a host/host configuration
    4c cooling in a dopant/host configuration

Fig 5 energy levels in optically cooled lasers
    a: spontaneous cooling
    b: cooling in a Raman laser
    c: stimulated cooling

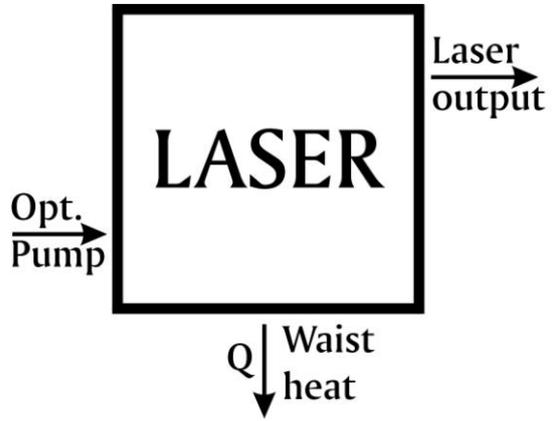 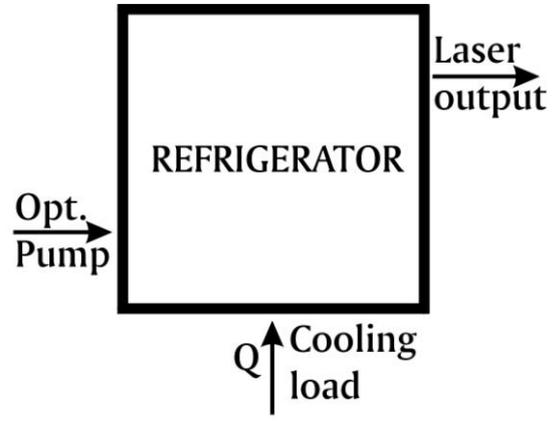

(1a)  (1b)

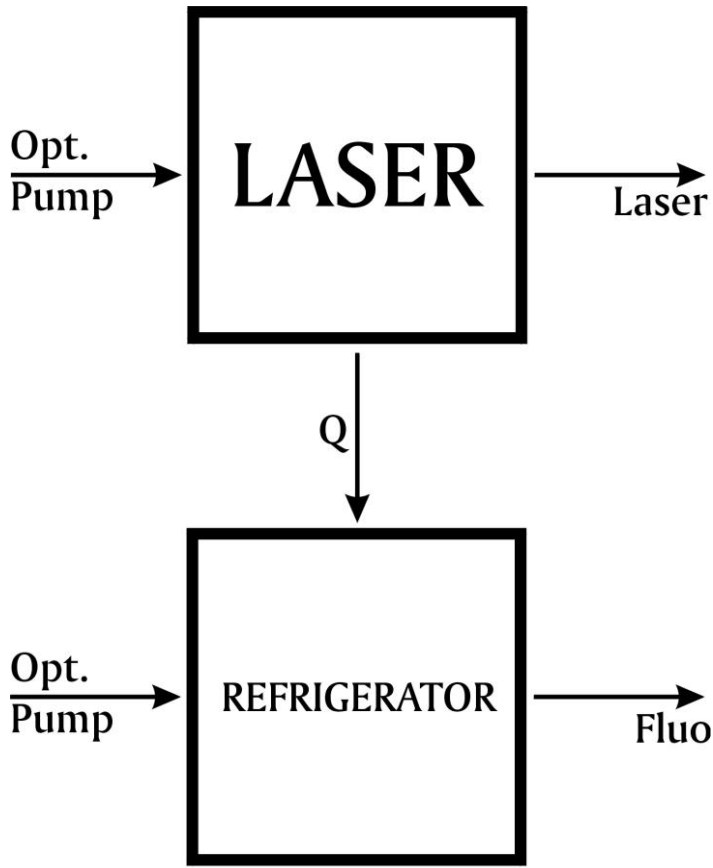

(2)

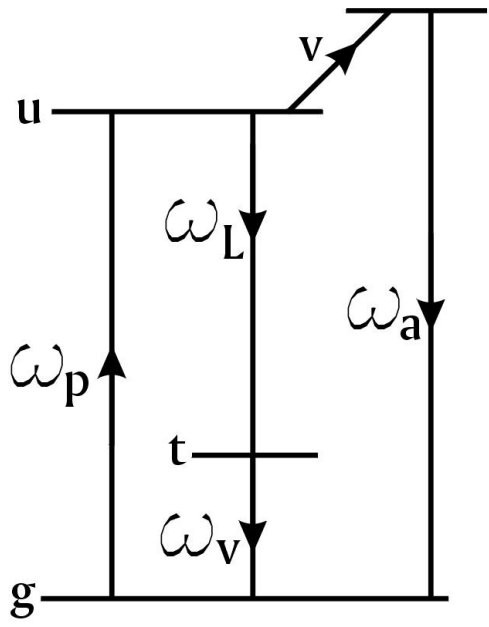 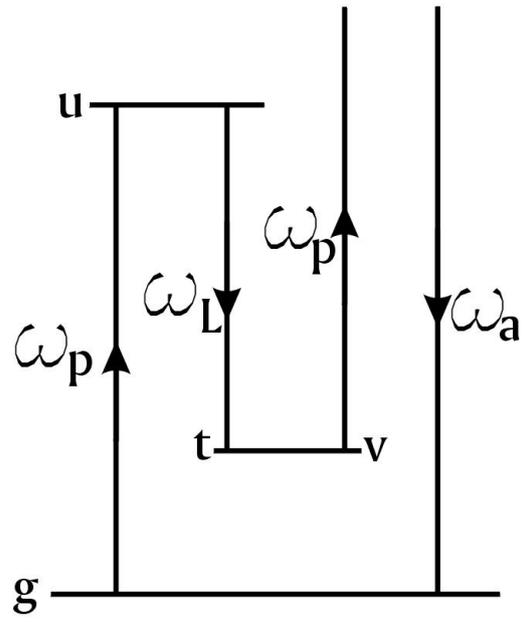

(3a)                                            (3b)

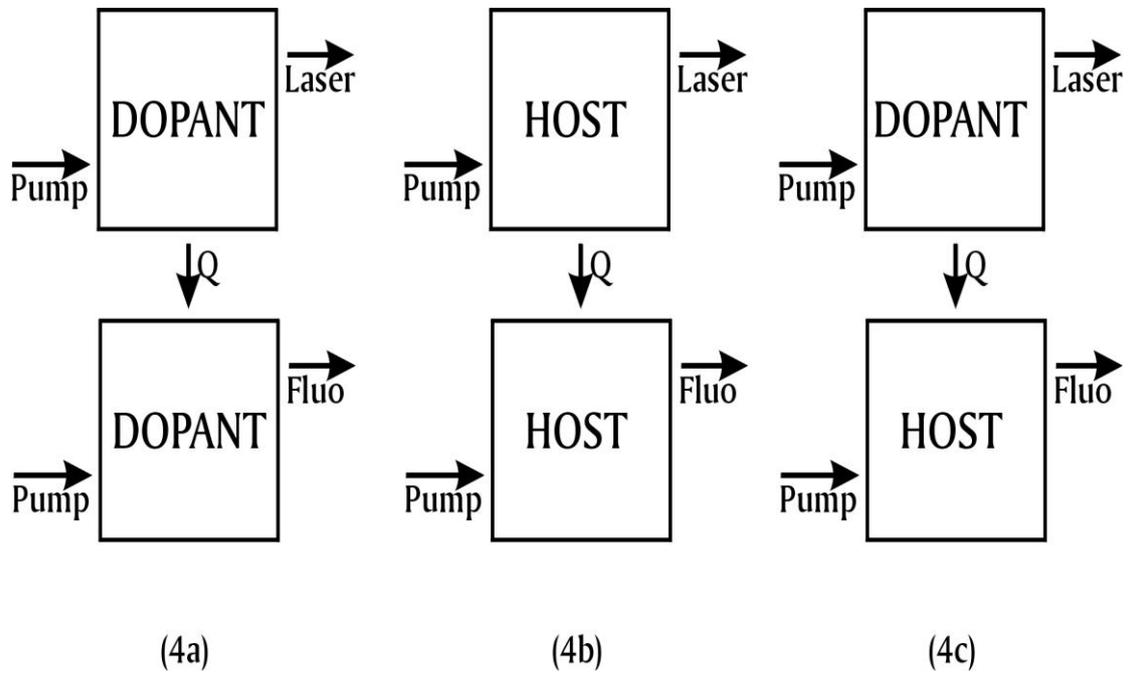

(4a)          (4b)          (4c)

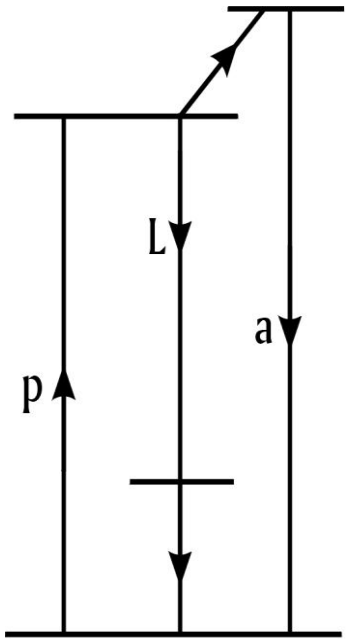 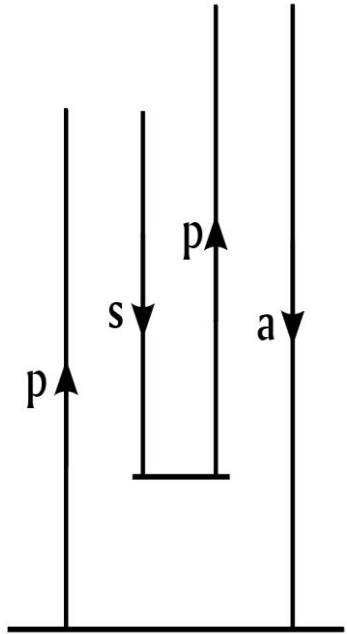 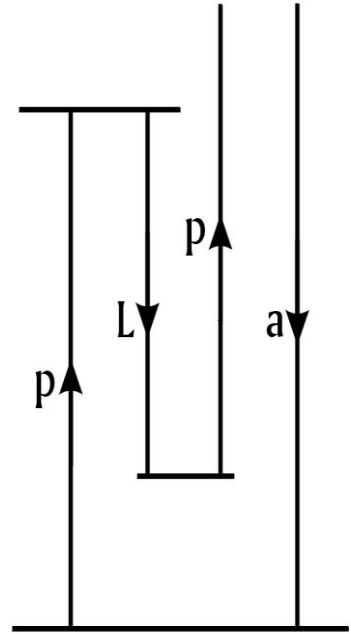

(5a)      (5b)      (5c)